APPLIED OPTICS

# Wafer-scale growth of large arrays of perovskite microplate crystals for functional electronics and optoelectronics

**Gongming Wang,[1,2]* Dehui Li,[1]* Hung-Chieh Cheng,[3] Yongjia Li,[3] Chih-Yen Chen,[1] Anxiang Yin,[1] Zipeng Zhao,[3] Zhaoyang Lin,[1] Hao Wu,[3] Qiyuan He,[1] Mengning Ding,[2,3] Yuan Liu,[3] Yu Huang,[2,3] Xiangfeng Duan[1,2]†**

**Methylammonium lead iodide perovskite has attracted intensive interest for its diverse optoelectronic applications. However, most studies to date have been limited to bulk thin films that are difficult to implement for integrated device arrays because of their incompatibility with typical lithography processes. We report the first patterned growth of regular arrays of perovskite microplate crystals for functional electronics and optoelectronics. We show that large arrays of lead iodide microplates can be grown from an aqueous solution through a seeded growth process and can be further intercalated with methylammonium iodide to produce perovskite crystals. Structural and optical characterizations demonstrate that the resulting materials display excellent crystalline quality and optical properties. We further show that perovskite crystals can be selectively grown on prepatterned electrode arrays to create independently addressable photodetector arrays and functional field effect transistors. The ability to grow perovskite microplates and to precisely place them at specific locations offers a new material platform for the fundamental investigation of the electronic and optical properties of perovskite materials and opens a pathway for integrated electronic and optoelectronic systems.**

## INTRODUCTION

Methylammonium lead iodide perovskite ($CH_3NH_3PbI_3$) has recently attracted intensive interest for its diverse optoelectronic applications, including solar cells (*1–11*), photodetectors (*12–17*), lasers (*18*, *19*), and light-emitting diodes (LEDs) (*20*). To date, most studies have been limited to bulk polycrystalline thin films made of perovskite materials (*21–27*). The ability to use lithography to pattern electronic materials is essential for integrated electronic and optoelectronic systems. Because they are soluble in various solvents (*28*), perovskite materials are not compatible with the typical lithographic process and thus cannot be easily patterned into discrete device arrays. This represents an urgent challenge to the advancement of this field. To this end, the direct growth of perovskite crystals at specific locations of prepatterned electrodes offers an alternative strategy for developing highly integrated systems (*29–31*). We report a facile two-step strategy for the scalable growth of large arrays of methylammonium lead iodide perovskite microplates with controlled physical dimensions and spatial location on diverse substrates (including silicon wafers, transparent glass, and prepatterned gold electrodes) and demonstrate that these directly grown microplate arrays can be used to create independently addressable photodetector arrays and well-performing field effect transistors (FETs) with the best reported field effect mobility to date. Our study presents the first successful patterned growth of perovskite crystals for integrated device arrays and solves the urgent challenges of the incompatibility of perovskite crystals with the typical lithographic process. The ability to grow perovskite crystals and to precisely place them at specific positions offers a new material platform for the fundamental investigation of the electronic and optical properties of perovskite materials and makes a necessary step toward integrated electronic and optoelectronic systems.

[1]Department of Chemistry and Biochemistry, University of California, Los Angeles, Los Angeles, CA 90095, USA. [2]California Nanosystems Institute, University of California, Los Angeles, Los Angeles, CA 90095, USA. [3]Department of Materials Science and Engineering, University of California, Los Angeles, Los Angeles, CA 90095, USA.
*These authors contributed equally to this work.
†Corresponding author. E-mail: xduan@chem.ucla.edu

## RESULTS

For the production of regular arrays of perovskite crystals, lead iodide ($PbI_2$) microplates were first grown on a substrate with prepatterned surface functionalization to control nucleation and growth and then converted into methylammonium lead iodide perovskites under methylammonium iodide ($CH_3NH_3I$) vapor. Figure 1A illustrates the schematic processing steps in producing patterned perovskite crystals (see Materials and Methods). The substrates (for example, $SiO_2$/Si) were first cleaned and functionalized with a self-assembled monolayer of (octadecyl)trichlorosilane (OTS) to produce a hydrophobic surface. A photolithography or electron beam lithography process, followed by oxygen plasma treatment, was used to create periodic arrays of hydrophilic areas on the hydrophobic surface.

The patterned substrate was treated with $PbI_2$ solution to produce $PbI_2$ seeds in hydrophilic regions (see Materials and Methods). Optical microscopy was used to monitor seeding and crystal growth. After the initial solution seeding process, small $PbI_2$ plates were clearly visible in the patterned hydrophilic area (Fig. 1B and fig. S1). In general, scanning electron microscopy (SEM) studies indicate that seeding crystals can be clearly visible in each hydrophilic square (fig. S1). The ability to selectively seed crystal nuclei in hydrophilic regions can be attributed to the periodic hydrophilic/hydrophobic patterns on the surface of the substrate. When an aqueous seeding solution flows over the surface of the substrate, a thin solution film can be trapped in hydrophilic regions to allow the nucleation of crystal seeds; on the other hand, there is no residue solution left in hydrophobic regions to produce crystal seeds. The seeded substrate was then immersed in saturated $PbI_2$ aqueous solution at 80°C to allow seeded growth of $PbI_2$ microplates. It is evident from the optical microscopy images that the $PbI_2$ plates became larger and larger with increasing growth time in $PbI_2$ source solution (Fig. 1, C and D), indicating that $PbI_2$ plate size can also be manipulated by varying growth time. The resulting $PbI_2$ plate arrays were finally intercalated with methylammonium iodide to form perovskite crystals in a home-built tube furnace system (see Materials and Methods).

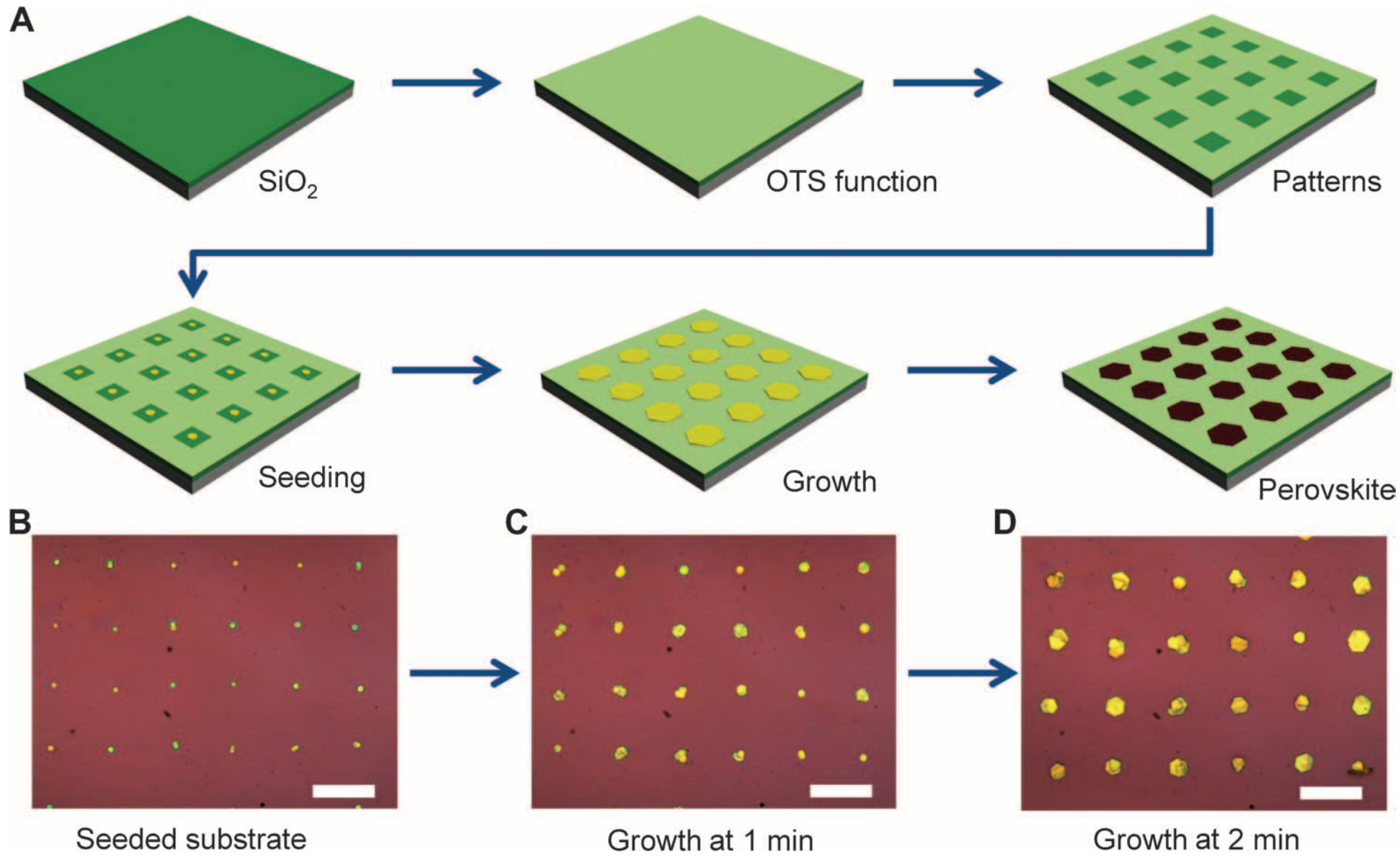

**Fig. 1. Schematic illustration of the patterned growth of regular arrays of perovskite microplate crystals.** (**A**) Schematic illustration of the procedure for preparing methylammonium lead iodide perovskite plates on a patterned substrate. The $SiO_2$/Si substrate was first functionalized with self-assembled monolayers of OTS to produce a hydrophobic surface and then lithographically patterned to create periodic arrays of hydrophilic areas. A hot aqueous $PbI_2$ solution (0.1 g/100 ml) was used as seeding solution to produce $PbI_2$ seed particles in hydrophilic regions using a flow seeding process. The seeded substrate was next immersed in a saturated $PbI_2$ solution at 80°C to further grow the seeds into larger $PbI_2$ microplates. Finally, $PbI_2$ microplate arrays were intercalated with methylammonium iodide vapor in a homebuilt tube furnace system. (**B** to **D**) Optical images of $PbI_2$ seed arrays after flow seeding process (B) and after further growth in saturated $PbI_2$ solution at 80°C for 1 min (C) and for 2 min (D). Scale bars, 40 μm.

## DISCUSSION

Successful control of crystal nucleation and growth makes it possible to grow large arrays of $PbI_2$ microplates with various arrangements over large areas of diverse substrates. Figure 2 (A to D) shows $PbI_2$ microplates grown on a $SiO_2$/Si substrate with hexagonal lattice patterns. The uniform dark-field optical microscopy image under low magnification indicates that patterned growth of $PbI_2$ can be achieved over large areas with high yield (Fig. 2A). Figure 2 (B and C) shows higher-magnification bright-field optical microscopy images of $PbI_2$ plates, displaying only one $PbI_2$ microplate in most of the patterned sites. Multinuclei formation is occasionally observed in a single-pattern site. The continued growth in such multinuclei sites could result in overlapped crystals with varied sizes or shapes. The SEM image of patterned crystal arrays further shows that most of the resulting $PbI_2$ microplates exhibit a smooth surface and lie flat on the surface of the substrate in predefined arrays (Fig. 2D). The size of most $PbI_2$ microplates is about 10 μm, and the thickness of the microplates is about 300 to 500 nm.

Our strategy for growing patterned $PbI_2$ microplate arrays is a general approach that can be readily used to grow crystal arrays with different lattice arrangements on diverse substrates. Figure 2 (E to H) shows the patterned growth of $PbI_2$ microplates in a square lattice pattern. Similar to growth in hexagonal lattice patterns, the growth of $PbI_2$ microplates in square lattice patterns is also highly uniform, demonstrating that our seed-mediated growth approach is very reliable and independent of pattern lattice arrangements. We have also grown $PbI_2$ microplates with different periodicities (including 20, 30, and 40 μm) (Fig. 2, I to K). The size of $PbI_2$ microplates apparently increases with increasing lattice spacing under the same growth conditions, which can be attributed to reduced competition from neighboring crystals and increased $PbI_2$ source supply for each crystal with increasing crystal-crystal distance (periodicity).

Successful control of nucleation and growth also enables us to grow $PbI_2$ microplates in nearly arbitrary patterns by using lithography to pattern preferential nucleation sites. For instance, by selectively patterning hydrophilic regions, we have successfully produced organized $PbI_2$ microplates to form "UCLA" characters, further demonstrating that our growth method is extraordinarily reliable and independent of pattern lattice shapes (Fig. 2L). To demonstrate the scalability of our approach, we have further shown that regular arrays of $PbI_2$ microplates can be readily produced across a 4-inch wafer with highly uniform and regular patterns (Fig. 2M). Besides $SiO_2$/Si wafer, we have also shown that patterned growth can be readily achieved on a transparent glass substrate (fig. S2).

We conducted x-ray diffraction (XRD) and transmission electron microscopy (TEM) studies to investigate the structural quality of the resulting $PbI_2$ crystals. Figure 2N shows the XRD pattern of $PbI_2$ microplate arrays grown on glass substrate. The XRD pattern only exhibits four dominant diffraction peaks, which can be assigned to the (001), (002), (003), and (004) facets of hexagonal $PbI_2$ (JCPDS number 07-0235) (*32*), demonstrating that the microcrystals are highly crystallized and well orientated on the substrate. A low-resolution TEM image shows that the $PbI_2$ crystal is hexagonal (Fig. 2O), consistent with SEM and optical microscopy studies. Electron diffraction of the microplate shows a single set of sixfold symmetric diffraction patterns (Fig. 2O, inset) along the [0001] zone axis, consistent with the hexagonal structure of $PbI_2$, and demonstrates that the prepared $PbI_2$ microplates are single crystals. A high-resolution TEM (HRTEM) image shows clear lattice fringes with *d*-spaces of the (110) and (−120) planes of hexagonal $PbI_2$ (Fig. 2P), further confirming the single crystalline structure. Together, these structural studies clearly demonstrate that we have achieved a well-controlled growth of $PbI_2$ microplate arrays with single crystal quality using a simple low-temperature seed-mediated solution growth method.

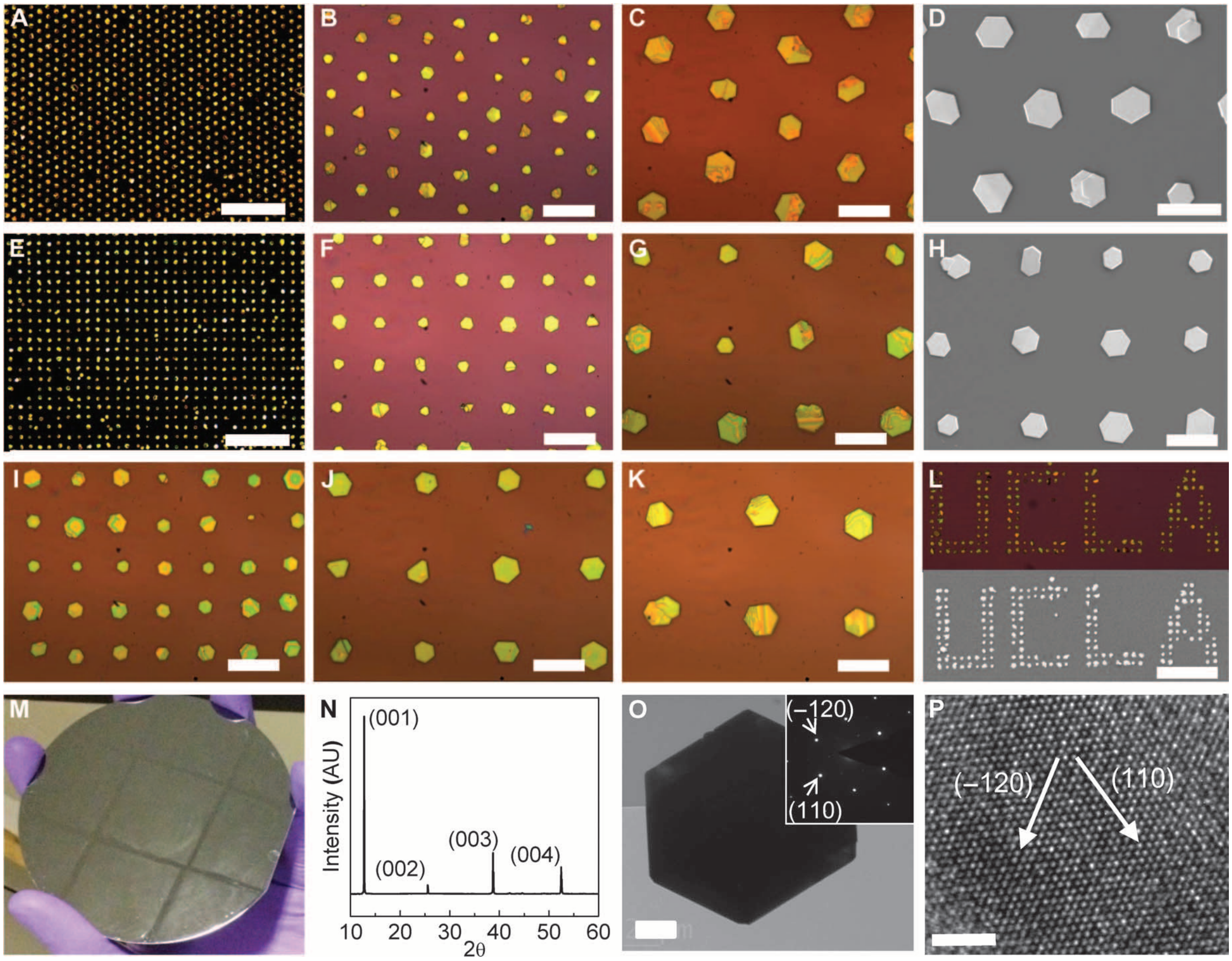

**Fig. 2. Growth of periodic arrays of $PbI_2$ microplate crystals.** (**A**) Dark-field optical microscopy image of $PbI_2$ microplates in hexagonal lattice patterns. Scale bar, 200 μm. (**B** and **C**) Higher-magnification bright-field optical microscopy images of $PbI_2$ microplates. Scale bars, 40 μm (B) and 20 μm (C). (**D**) SEM image of $PbI_2$ microplates in a hexagonal lattice pattern. Scale bar, 20 μm. (**E**) Dark-field optical microscopy image of $PbI_2$ plates in square lattice patterns (8 μm × 25 μm). Scale bar, 200 μm. (**F** and **G**) Higher-magnification bright-field optical microscopy images of $PbI_2$ microplate arrays. Scale bars, 40 μm (F) and 20 μm (G). (**H**) SEM image of $PbI_2$ plates in a square lattice pattern. Scale bar, 20 μm. (H to **K**) Bright-field optical microscopy images of $PbI_2$ microplates in square lattice patterns with periodicities of 20, 30, and 40 μm, respectively. Scale bars, 20 μm. (**L**) Dark-field optical microscopy image (top) and SEM image (bottom) of $PbI_2$ microplates arranged in "UCLA" patterns. Scale bar, 100 μm. (**M**) Digital photo of patterned growth on a 4-inch silicon wafer. (**N**) XRD pattern of $PbI_2$ microplate arrays on a glass substrate. (**O**) Low-resolution TEM image of a $PbI_2$ microplate. Scale bar, 2 μm. (Inset) Electron diffraction pattern. (**P**) HRTEM image of the $PbI_2$ microplate with lattice fringes clearly resolved. Scale bar, 2 nm.

The prepared $PbI_2$ microplates can be further converted into methylammonium lead iodide perovskites through a gas-solid heterophase intercalation process under methylammonium iodide vapor in a homebuilt tube furnace system (see Materials and Methods). Methylammonium iodide precursor was synthesized using a reported solution method and recrystallized in diethyl ether/methanol (*33*). Compared to the conventional liquid-solid conversion in organic solvents, gas-solid intercalation could prevent the chemical dissolution of $PbI_2$ and perovskite crystals in organic solvents and retain the morphology and crystalline quality of the perovskite microplates well (*32*).

Figure 3A shows the crystal structures of $PbI_2$ and methylammonium lead iodide perovskite. Before intercalation, hexagonal $PbI_2$ exhibits a layered structure where each octahedron shares two equatorial halide atoms and one axial halide atom with its neighbors. During intercalation, the methylammonium group is inserted into the center of eight octahedrons and relocates the equatorial halide atoms, inducing a twisting of the $PbI_2$ octahedrons (*32*). The converted methylammonium lead iodide perovskites have a tetragonal structure, and the octahedrons in perovskites only share one halide atom with their neighbors in both equatorial and axial directions. Optical microscopy, SEM, and TEM images of the converted perovskites show a similar hexagonal plate with a clean surface (Fig. 3, B to D and F), suggesting that the conversion of $PbI_2$ into perovskite does not significantly change the overall morphology of the crystals.

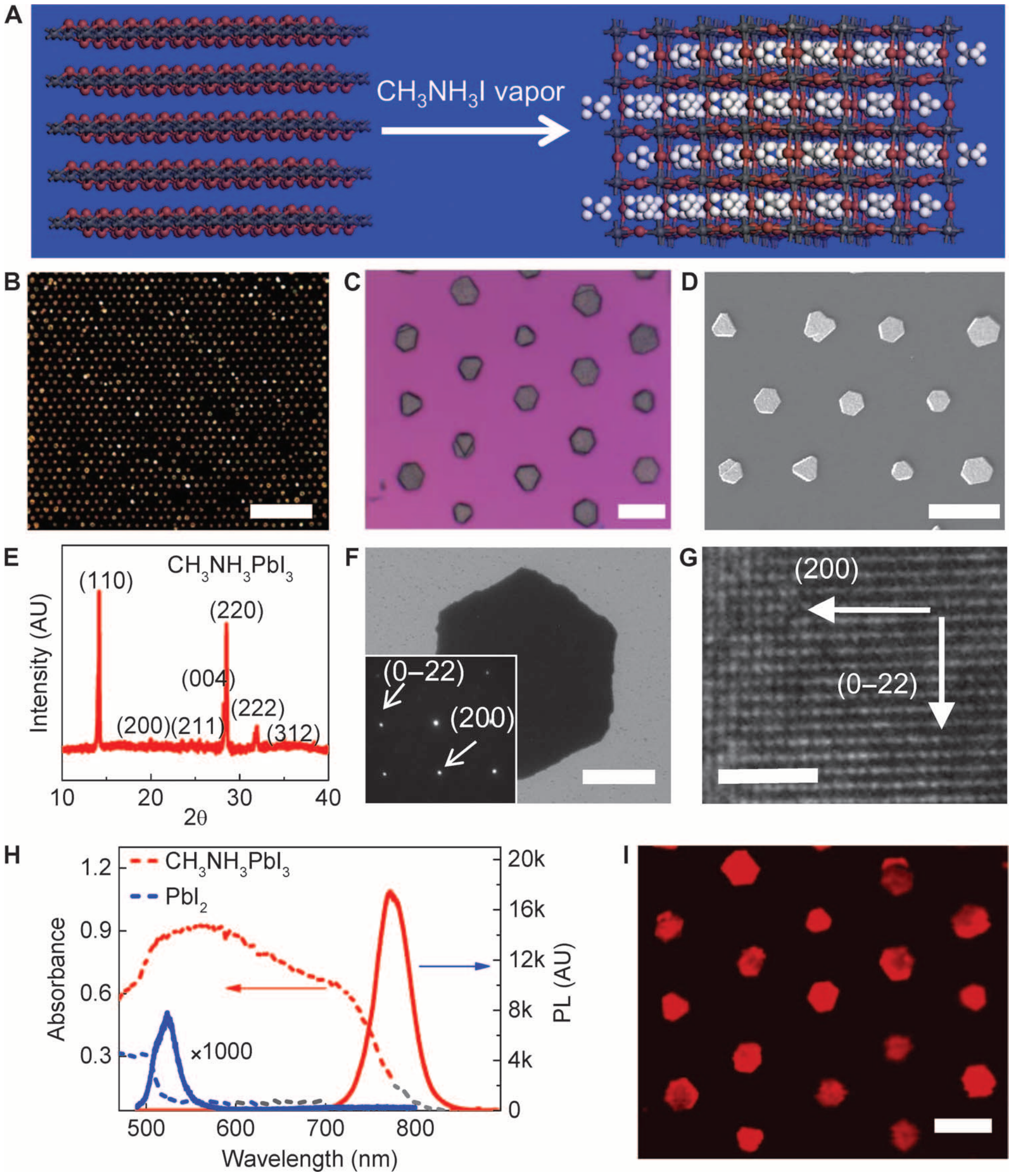

**Fig. 3. Conversion of $PbI_2$ microplates into high-quality perovskite crystals.** (**A**) Schematic illustration of the change in lattice structure from layered $PbI_2$ to tetragonal perovskite after methylammonium iodide intercalation. (**B**) Dark-field optical microscopy image of perovskite microplate arrays. Scale bar, 200 μm. (**C**) Higher-magnification bright-field optical microscopy image of perovskite microplate arrays. Scale bar, 20 μm. (**D**) SEM image of perovskite arrays. Scale bar, 20 μm. (**E**) XRD pattern of converted perovskite microplates on a glass substrate. (**F**) Low-resolution TEM image of a perovskite microplate. Scale bar, 2 μm. (Inset) Electron diffraction pattern of the microplate. (**G**) HRTEM image of a perovskite microplate with clearly resolved lattice fringes. Scale bar, 2 nm. (**H**) UV-vis absorption and photoluminescence (PL) spectra of $PbI_2$ and perovskite microplates. (**I**) Spatially resolved mapping image of the photoluminescence of a perovskite crystal array. Scale bar, 20 μm.

XRD studies of converted crystals have shown that all of the diffraction peaks could be indexed to the tetragonal structure of perovskite materials and that the diffraction peaks of $PbI_2$ disappeared entirely (Fig. 3E), suggesting a complete conversion of the layered $PbI_2$ into perovskite crystals (*7*, *32*). The electron diffraction pattern of a converted perovskite microplate shows a single set of diffraction spots that can be indexed to the tetragonal structure of the perovskite crystals (Fig. 3F, inset), demonstrating the excellent crystalline quality of converted perovskite microplates. An HRTEM image shows clear lattice fringes with *d*-spaces of the (200) and (0–22) planes of the perovskite structure (Fig. 3G), further confirming the excellent crystallinity quality of the resulting perovskite microplates. The perovskite crystals are not stable under electron beam, and the lattice fringes blur with increasing electron beam irradiation. In addition, energy-dispersive x-ray (EDX) analysis reveals that the atomic ratio of Pb to I is about 1:2.9 (fig. S3), further confirming the successful conversion into perovskite crystals through the vapor intercalation method.

We have further characterized the optical properties of converted perovskite microplates. Before conversion, $PbI_2$ crystals exhibit an absorption edge at about 525 nm (Fig. 3H, blue dashed line) and a photoluminescence peak at 522 nm (Fig. 3H, blue solid line). In comparison, converted perovskite crystals exhibit an absorption edge at about 800 nm (Fig. 3H, red dashed line), consistent with literature-reported values of perovskites prepared through both solution and vapor conversion methods (*3*, *32*). Furthermore, converted perovskite crystals exhibit a very strong room temperature photoluminescence peak at about 770 nm (Fig. 3H, red solid line), consistent with previous studies (*18*). Photoluminescence in perovskite crystals is more than three orders of magnitude stronger than that in $PbI_2$ crystals, indicating the high crystalline quality and excellent photoluminescence efficiency of the resulting perovskite crystals. Photoluminescence mapping of the prepared perovskite crystal array further shows strong photoluminescence emission from the entire perovskite crystal array (Fig. 3I). Together, our studies clearly demonstrate that $PbI_2$ microplates have been successfully converted into perovskite crystals with excellent crystalline quality and optical properties.

Control of crystal nucleation and growth on predefined sites allows us to directly grow perovskite crystals onto prepatterned electrodes to form functional devices. This direct growth method readily enables us to fabricate discrete devices without using photolithography or electron beam lithography processes that could damage the perovskite crystals and uniquely solves the urgent challenges of the incompatibility of perovskite crystals with the typical lithographic process. With this approach, large arrays of independently addressable devices can be readily prepared by selective nucleation and growth of perovskite crystals between arrays of electrode pairs. On the basis of the growth mechanism discussed previously, microplate crystals first nucleate in hydrophilic sites between two electrodes on the prepatterned substrate and gradually extend to the electrodes as the microplates grow larger with time. As a result, a large array of two-probe devices can be obtained, with each pair of probe electrodes bridged by a singlet perovskite microplate crystal (Fig. 4A). Electrical measurements of these two-probe devices (channel length, 8 μm) show a nearly zero dark current and a nearly linear current-voltage (*I-V*) behavior under illumination (Fig. 4B), indicating excellent photoresponse of the perovskite crystals. The photocurrent–to–dark current ratio of our devices can reach up to three orders of magnitude. Responsivity, defined as the ratio of photocurrent to incident light power, was calculated to be ~7 A/W, with a corresponding photocurrent gain of about 18. The response speed of our devices is characterized by a rising time and a falling time of ~500 μs (Fig. 4C, inset), which is limited by our measurement capability. Photocurrent increases with increasing incident light power and shows a sublinear dependence on light power (Fig. 4C). The sublinear dependence can be ascribed to the complex photogenerated carrier dynamics in semiconductors (*34*), which has been observed in a number of semiconductor photodetectors. The performance of our photodetectors can be further improved by optimizing the fabrication process and/or device configuration. For example, by reducing channel length to ~100 nm or so, a significantly higher photoresponsivity (~40 A/W) and a photocurrent gain of ~100 can be achieved. Although the performance of these photoconductance-based lateral photodetectors is not as good as that of the recently reported photodiode-based photodetectors (*12*, *15*), the photoresponsivity of our device is at least three orders of magnitude higher than that of perovskite nanowire mesh photodetectors (~5 mA/W) with the same photoconductance sensing mechanism (*17*).

Overall, our two-terminal devices show a high degree of reproducibility and yield. We have measured more than 40 devices, all of which exhibit a strong photoresponse under illumination. In general, as long as microplate crystals are in contact with both electrodes, the devices will show an obvious photoresponse with essentially a unit device yield. With precise control of microcrystal location and high device yield, it is possible to create large photodetector arrays. To this end, we created a large-scale (10 × 10) photodetector array on a transparent glass substrate (fig. S4) and placed a "U"-shaped mask on the back of the glass substrate so that the incident light can only reach the area in the transparent U-shaped region (Fig. 4D). Measurement of the photoresponse of the entire array indicates that almost all of the devices exposed to the U-shaped light illumination show a clear photocurrent response with a 92% device yield (Fig. 4E). Three devices did not respond because of the absence of perovskite crystals between the electrodes. A spatially resolved map of the photocurrent amplitude generated by the photodetector array clearly shows a U-shaped photoresponse area (Fig. 4F), demonstrating that such arrays of two-terminal devices can function as effective photoimaging arrays. These studies demonstrate an important step toward integrated device applications based on individual perovskite crystals. We noted some variations in the photocurrent amplitude from each pixel, which may be attributed to differences in size, in the orientation of perovskite crystals, and in the contact of perovskite crystals with gold electrodes, and may be further improved upon future optimization of nucleation and growth.

Controlled growth of perovskite crystals on predefined electrodes also enables us to probe the intrinsic charge transport properties of perovskite materials, which remain elusive in bulk polycrystalline thin films because of variability in the spin-coating process and extensive grain boundary scattering and trapping. To this end, we have conducted systematic temperature-dependent studies to investigate the fundamental electrical transport properties of our perovskite microplate crystals. The perovskite crystal is explored as the semiconducting channel of a FET on a $SiO_2$/Si substrate, with two gold electrodes acting as source and drain electrodes (channel length, 40 μm) and with the $SiO_2$/Si substrate acting as gate dielectrics and gate electrode. A set of representative output curves (source drain current $I_{sd}$ versus source drain voltage $V_{sd}$) of a perovskite microplate crystal FET under various gate voltages ($V_g$) shows typical transistor characteristics (Fig. 4G). A slightly nonlinear $I_{sd}$-$V_{sd}$ behavior is observed near zero bias, indicating that contact resistance is not yet fully optimized. The transfer characteristics show dominant n-type behavior, with conductance increasing with increasing positive gate voltage (Fig. 4H). Slight p-type behavior is also observed at high negative gate voltage, indicating ambipolar characteristics. Similar to what is commonly seen in the literature, our devices also show considerable hysteresis, which has been attributed to field-induced ion drift, ferroelectric, and/or trap state filling effects on perovskites (*35*, *36*). The maximum on/off ratio ($I_{on}/I_{off}$) is nearly six orders of magnitude, which is better than that of recently reported polycrystalline thin film perovskite transistors (*35*, *36*). Furthermore, field effect carrier mobility can also be extracted from the transfer characteristics. The existence of hysteresis may introduce errors in mobility determination, with the forward gate sweep underestimating field effect mobility and with the backward gate sweep overestimating field effect mobility. To properly evaluate carrier mobility in our microplate crystals, we have extracted carrier mobility values based on both forward sweep and backward sweep transfer characteristics. Field effect electron mobility can reach up to ~2.5 $cm^2$/V·s (backward sweep) and ~1 $cm^2$/V·s (forward sweep) at 77 K, both of which are significantly better than those obtained in spin-coated polycrystalline perovskite thin films (<$10^{-1}$ $cm^2$/V·s at 77 K) under a similar measurement configuration (*35*). We have also performed a statistical analysis of the distribution of mobility values extracted from independent measurements on 27 devices (Fig. 4I). Although the absolute values of electron mobility display some variability, they are all consistently higher than those recently reported in spin-coated polycrystalline thin films (*35*). These results clearly demonstrate the high quality and unique advantages of perovskite microplate crystals compared with typical polycrystalline thin films. Our device is not yet fully optimized, and there is an apparent contact barrier, as indicated by the nonlinear $I_{sd}$-$V_{sd}$ behavior near zero bias. Therefore, we believe that the intrinsic carrier mobility of perovskite microplate crystals could be further improved upon optimization of contact resistance.

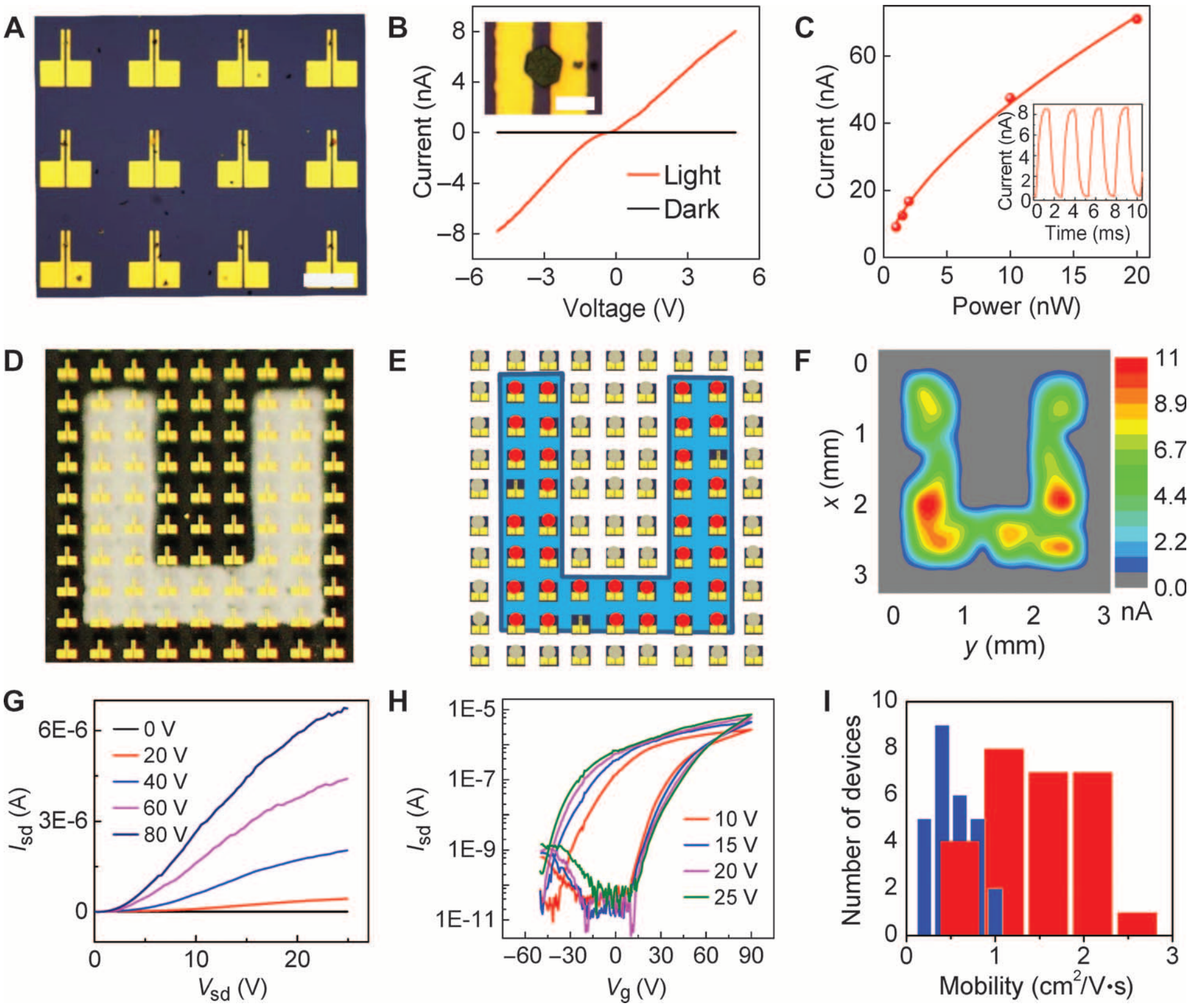

**Fig. 4. Selective growth of perovskite crystals on prepatterned electrodes for photodetector arrays and FETs.** (**A**) Optical microscopy image of perovskite microplate crystals bridging prepatterned arrays of electrode pairs. Scale bar, 200 μm. (**B**) Typical *I-V* curve of a perovskite crystal under dark and light illumination. (Inset) Optical microscopy image of a perovskite crystal bridging two gold electrodes. The distance between the electrode pair is 8 μm. Scale bar, 20 μm. (**C**) Photocurrent versus incident light power of a typical device (source-drain voltage, 5 V). (Inset) Response speed of a typical two-probe device (laser illumination, 488 nm; power, 1.2 nW; source-drain voltage, 5 V). (**D**) Optical image of photodetector arrays with a U-shaped mask (U-shaped transparent area). (**E**) Schematic illustration of the device arrangement of photodetector arrays with a U-shaped mask under blue LED illumination (wavelength, 463 nm; power density, 600 μW/cm$^2$). Red dots, devices with photocurrent; gray dots, devices without photocurrent. (**F**) Photocurrent mapping of photodetector arrays with a U-shaped mask under blue LED illumination. (**G** and **H**) Output ($V_g$ = 0, 20, 40, 60, and 80 V) (G) and transfer ($V_{sd}$ = 10, 15, 20, and 25 V) (H) characteristics of FET based on a perovskite microplate crystal at 77 K. (**I**) Distribution of the field effect electron mobility of 27 perovskite microplate transistors measured at 77 K. Red, mobility derived from backward sweep; blue, mobility derived from forward sweep.

In summary, we have reported a facile low-temperature solution method for the scalable growth of perovskite microplate crystals with controlled spatial location and periodic arrangement over a large area (up to 4-inch wafer). We further show that this approach can enable the selective growth of perovskite crystals on prepatterned gold electrodes to create independently addressable photodetector arrays and discrete transistors. Our study presents the first successful patterned growth of large arrays of perovskite crystals—a critical advancement in controlled production of patterned perovskite crystals for independently addressable electronic and optoelectronic device arrays that has been very difficult to achieve with other conventional lithographic techniques. The direct growth of patterned arrays timely solves the urgent challenges of the incompatibility of perovskite crystals with the typical lithographic process. The entire process of crystal growth and device fabrication occurs at a rather low temperature (<120°C) and can be readily applied for the growth of functional perovskite device arrays on other low-cost substrates, including plastics, for large-area flexible electronic and optoelectronic applications. The controlled growth of perovskite crystals on predefined patterns is an essential step toward the development of perovskite materials. Recent studies have shown that improved crystalline quality in perovskite materials is beneficial to device performance (*37*–*39*). Our studies demonstrate that high-quality perovskite FET devices can deliver an electron mobility value higher than 1.0 cm$^2$/V·s at 77 K, far exceeding those achieved in spin-coated polycrystalline thin films. The ability to grow high-quality perovskite crystals with controlled physical dimensions and to precisely place them at specific locations can thus provide a powerful new material platform for the fundamental investigation of the intrinsic electronic and optoelectronic properties of perovskite materials and can open up exciting opportunities to explore perovskite arrays for diverse electronic and optoelectronic systems such as transistors, photodetectors, solar cells, LEDs, or laser diodes.

## MATERIALS AND METHODS

### Preparation of patterned substrate

Precleaned $SiO_2$/Si wafer was dipped into a mixture of hexane and OTS solvent (500:1) for 10 min and rinsed with acetone for 30 s to obtain self-assembled monolayers of hydrophobic OTS on the surface. A lithography process, followed by oxygen plasma treatment, was used to selectively remove OTS from designated locations to create periodic arrays of hydrophilic regions. Finally, lithography resist was removed by dipping the substrate into acetone for a few minutes. After these treatments, the substrate exhibited hydrophobic properties overall but periodically showed hydrophilic properties at selectively defined sites. The freshly prepared patterned substrates were further used for the seeding process.

### Patterned growth of $PbI_2$ microplate arrays

$PbI_2$ seed arrays were achieved using a solution flow process. A dilute $PbI_2$ aqueous solution (0.1 g/100 ml) prepared at 80°C was dropped on the tilted patterned $SiO_2$/Si substrate. $PbI_2$ seeds are generated in each hydrophilic region when the solution flows through the surface of the substrate. The seeded substrate was then placed into a saturated $PbI_2$ aqueous solution at 80°C (~0.3 g/100 ml) for several minutes to allow the $PbI_2$ seeds to grow larger. Before immersion into the hot $PbI_2$ precursor solution, the seeded substrate was first prewarmed by the vapor on top of the solution for 30 s to avoid rapid nucleation on the substrate as a result of a large difference in temperature.

### Conversion of $PbI_2$ microplates into perovskite crystals

Methylammonium iodide powder was synthesized using a previously reported solution method (*33*) and used as intercalation source. The methylammonium iodide source was placed at the center of a 1-inch tube furnace (30760-058; Thermo Scientific), and the substrates with $PbI_2$ microplate arrays were placed 5 to 6 cm downstream. Before conversion, the tube furnace was vacuumed and refilled with argon at least three times to completely remove the air in the quartz tube. Conversion was conducted at a pressure of 100 mbar, with argon (100 standard cubic centimeters per minute) used as carrier gas for several hours. The actual temperature at the methylammonium iodide source region was 140°C, and that at the $PbI_2$ microplate substrate region (measured by a thermocouple probe) was 120°C. Finally, the tube furnace was naturally cooled down to room temperature.

### Material characterization

The structure of $PbI_2$ microplate arrays was characterized using an optical microscope (Olympus), a scanning electron microscope (JSM-6700F FE-SEM; JEOL), and an x-ray diffractometer (PANalytical X'Pert Pro). Photoluminescence was measured on a confocal micro-Raman system (LabHR; HORIBA) in backscattering configuration excited by an argon-ion laser (488 nm) with 1.5 μW of excitation power. Photoresponse was measured on the same micro-Raman system coupled with a computer-controlled analog-to-digital converter (model 6030E; National Instruments). The photoresponse of the U-shaped array was measured in a probe station (TTP4; LakeShore) coupled with a computer-controlled analog-to-digital converter (model 6030E; National Instruments) and illuminated by a blue LED with a peak wavelength of 463 nm and a power density of 600 $\mu W/cm^2$. For measurement of response speed, a mechanical chopper was used to modulate the incident light. FET device measurements were performed in a probe station (TTP4; LakeShore) coupled with a precision source/measurement unit (B2902A; Agilent Technologies). The scanning rate for transport measurement was 20 V/s, and the devices were prebiased at the opposite voltage for 30 s before each measurement.

## SUPPLEMENTARY MATERIALS

Supplementary material for this article is available at http://advances.sciencemag.org/cgi/content/full/1/9/e1500613/DC1

Fig. S1. SEM characterization of seeded substrate using a flow seeding process.
Fig. S2. Optical microscopy image of $PbI_2$ plates grown on a transparent glass substrate.
Fig. S3. EDX studies of prepared perovskite plates.
Fig. S4. Digital image of perovskite photodetector arrays on a transparent glass substrate.

**Funding:** We acknowledge the support of the Electron Imaging Center for Nanomachines at University of California Los Angeles, which was funded by the NIH–National Center for Research Resources (grant CJX1-443835-WS-29646) and the NSF Major Research Instrumentation Program (grant CHE-0722519). We also acknowledge the Nanoelectronics Research Facility for providing technical support. X.D. was supported by the U.S. Department of Energy, Office of Basic Energy Sciences, Division of Materials Science and Engineering (award DE-SC0008055). Y.H. was supported by the NSF (grant EFRI-1433541). **Author contributions:** X.D. and Y.H. designed the experiments and supervised the research. G.W. and D.L. performed the experiments. All of the other authors contributed to material characterization. All authors discussed the results and commented on the manuscript. **Competing interests:** The authors declare that they have no competing interests. **Data and materials availability:** Data will be made available upon request by emailing xduan@chem.ucla.edu.